\documentstyle[12pt]{article}

\def\De{{\Delta}}
\def\Ga{{\Gamma}}
\def\Ps{{\Psi}}
\def\al{{\alpha}}
\def\be{{\beta}}
\def\bfB{{\bf B}}
\def\bfI{{\bf I}}
\def\bfP{{\bf P}}
\def\bff{{\bf f}}
\def\bfj{{\bf j}}
\def\bfk{{\bf k}}
\def\bfv{{\bf v}}
\def\bfx{{\bf x}}
\def\bfz{{\bf z}}
\def\cald{{\cal D}}
\def\calo{{\cal O}}
\def\cals{{\cal S}}
\def\const{{\rm const}}
\def\de{{\delta}}
\def\et{{\eta}}
\def\ga{{\gamma}}
\def\la{{\lambda}}
\def\om{{\omega}}
\def\ps{{\psi}}
\def\rh{{\rho}}
\def\smax{{\mathop{\scriptstyle\max}}}
\def\ta{{\tau}}
\def\th{{\theta}} 
\newcommand{\Alfvenic}{{\rm Alfv\'enic}}             
\newcommand{\Alfven}{{\rm Alfv\'en}}                 
\newcommand{\bfal}{{\mbox{\boldmath $\bf\alpha$}}}
\newcommand{\bfna}{{\mbox{\boldmath $\bf\nabla$}}}
\newcommand{\bfps}{{\mbox{\boldmath $\bf\psi$}}}
\newcommand{\bqy}{\begin{eqnarray}}
\newcommand{\bq}{\begin{equation}}
\newcommand{\eqy}{\end{eqnarray}}
\newcommand{\eq}{\end{equation}}
\newcommand{\p}{{\partial}}
\newcommand{\wh}[1]{{\widehat{#1}}}
\newcommand{\wt}[1]{{\widetilde{#1}}}
\input psfig
 
\def\thesection       {\Roman{section}}

\begin{document}
\bibliographystyle{../macros/ieeetr} 

\baselineskip18pt  
 
\title{Iso-topological relaxation, coherent structures,
and Gaussian turbulence in two dimensional magnetohydrodynamics} 

\author{
  M.~B. Isichenko\thanks{
    Institute for Fusion Studies, The University of Texas at Austin,
    Austin, TX 78712.  
    Present address:  Fusion Research Center, The University of Texas at 
    Austin, Austin, TX 78712.  
    }{ }
  and
  A.~V. Gruzinov\thanks{
    Department of Physics 0319, University of California, San Diego, 
    La Jolla, CA 92093.
    }
  }
\date{January 1994}

\maketitle   
 
\begin{abstract}
 
The long-time relaxation of ideal two dimensional (2D)
magnetohydrodynamic (MHD) turbulence subject to the conservation of
two infinite families of constants of motion---the magnetic and the
``cross'' topology invariants---is examined.  The analysis of the Gibbs
ensemble, where all integrals of motion are respected, predicts the
initial state to evolve into an equilibrium, stable coherent structure
(the most probable state) and decaying Gaussian turbulence
(fluctuations) with a vanishing, but always positive temperature.  The
non-dissipative turbulence decay is accompanied by decrease in both the
amplitude and the length scale of the fluctuations, so that the
fluctuation energy remains finite.  The coherent structure represents a
set of singular magnetic islands with plasma flow whose magnetic
topology is identical to that of the initial state, while the energy
and the cross topology invariants are shared between the coherent
structure and the Gaussian turbulence.  These conservation laws suggest
the variational principle of iso-topological relaxation which allows us
to predict the appearance of the final state from a given initial
state.  For a generic initial condition having $x$ points in the magnetic
field, the coherent structure has universal types of singularities:
current sheets terminating at $Y$ points.  These structures, which are
similar to those resulting from the 2D relaxation of magnetic field
frozen into an ideally conducting viscous fluid, are observed in the
numerical experiment of D.~Biskamp and H.~Welter [Phys.\ Fluids {\bf B
1}, 1964 (1989)] and are likely to form during the nonlinear stage of
the kink tearing mode in tokamaks.  
\label{loc1}  
The Gibbs ensemble method developed in this work admits extension to
other Hamiltonian systems with invariants not higher than quadratic in
the highest-order-derivative variables.  The turbulence in two
dimensional Euler fluid is of a different nature: there the coherent
structures are also formed, but the fluctuations about these
structures are non-Gaussian.

\end{abstract} 

52.35.Ra, 47.25.Cg.

\thispagestyle{empty}
\clearpage
\tableofcontents
\thispagestyle{empty}
\clearpage\addtocounter{page}{-2}
 
\section{Introduction}
\label{sec:introduction} 

Turbulence in two dimensional fluids is simple enough to allow some
analytically tractable models
\cite{Onsager49,KM80,Kuzmin82,Polyakov92} and high-resolution
computation \cite{BW89,MSMOM91,CMcWPWY91,CMcWPWY92,MMSMO92}, but still
sufficiently complex to exhibit the general challenging nature of
turbulence.  In addition, many features of plasma and geophysical
fluid dynamics are essentially two dimensional.
 
The principal difficulty encountered by any analytical approach to
fluid turbulence is that the underlying equations of motion are
nonintegrable, a quality independent of the abilities of the
researcher.  To overcome this difficulty, two major analytical
approaches have been attempted so far, namely, (a)~statistical
closures dealing with simplifying, but usually uncontrollable,
modifications of the dynamical equations and (b)~considerations of
{\em a priori\/} equally probable states constrained by the known
integrals of motion.  In this work we take the second approach, where
the consistency requirement is to allow for {\em all\/} constants of
the motion.
 
The conventional Boltzmann-Gibbs statistical mechanics, when applied to
partial differential equations \cite {Tasso87,LRS88,LRS89,Pomeau92a},
encounters a fundamental obstacle because of the underlying infinite
number of degrees of freedom and the necessity to use a finite ($N$)
dimensional approximation.  The continuum limit $N\to\infty$, in common
to all classical fields, results in the so-called ultraviolet
catastrophe (the divergence of energy at finite temperature), a problem
which goes back to Jeans.  Unlike the equilibrium electromagnetic
radiation, the ultraviolet catastrophe in fluid turbulence cannot be
remedied by quantization and should be resolved within the classical
framework.  The tendency toward the equipartition of energy between the
degrees of freedom (by $T/2$ for each degree, where $T$ is the energy
temperature) in a closed continuum system can only be satisfied by
letting the temperature to zero.  In fact, the way the temperature goes
to zero as the number of degrees of freedom goes to infinity is the
heart of the problem.  Nontrivial equilibrium states are obtained when
there are more than one integrals of motion, which diverge at different
rates as $N\to\infty$.
 
The absence of a well-defined concept of measure in a functional
(infinite dimensional) space requires an $N$ dimensional
discretization, even though the final results are obtained by letting
$N\to\infty$.  
\label{loc2}
Lee \cite {Lee52} was the first to use the truncated Fourier series
to show the validity of an infinite dimensional Liouville theorem.  The
choice of the discrete variables is not unique, and one ought to make sure
that the results of the statistical theory of turbulence be invariant
with respect to the way this choice is made.

A less formal, physical motivation for the discretization procedure can
be found in the finiteness of the number $N(t)$ of the ``effectively
excited'' collective modes at any finite time $t$.  Under ``effectively
excited'' we mean modes with amplitudes not yet exponentially small (a
smooth field, for instance, has exponentially small high Fourier
harmonics).  There are examples of $N(t)$ becoming infinite in finite
time \cite {FS91}, even when a real-space collapse \cite{Zakharov72}
does not occur.  It appears, however, that in two dimensional ideal
hydrodynamics and magnetohydrodynamics $N(t)$ behaves algebraically,
$N(t)\simeq(t/\ta_A)^p$, so that the time of the doubling of $N(t)$ is
of order $t$ and the time of the increment of $N(t)$ by one is of order
$\de t=\ta_A(t/\ta_A)^{1-p}$, if $0<p<1$.  Hence, on time scale longer
than the eddy turnover (nonlinear mode interaction) time $\ta_A$, one
may expect an equilibrium statistical distribution among the
effectively excited $N(t)$ modes.  By letting $N(t)\gg1$ in this
distribution, we shall infer the most probable direction of the
long-time system evolution.  The probability of a deviation from the
most probable state goes to zero as the number of degrees of freedom
goes to infinity.  As for a continuum conservative system we really
mean $N(t)\to\infty$ as $t\to\infty$, the probabilistic nature of the
statistical prediction assumes a rather deterministic quality.

Artificial finite dimensional approximations are notorious for
destroying an infinity of topological invariants (also known as
freezing-in integrals or Casimirs), which impose important constraints
on the evolution.  An interesting alternative proposed by Zeitlin \cite
{Zeitlin91}, whereby an $N$ dimensional hydrodynamic-type system
conserves $\sim\sqrt{N}$ invariants, is not quite suitable for
continuum fields because of the implied periodicity in the Fourier
space corresponding to modulated point vortices in real space.  It
would be very interesting to construct other ``meaningful'' (that is,
having many invariants) finite-mode hydrodynamics with well-behaved
real-space velocity fields.
 
So far, most statistical theories of continuum hydrodynamics,
\label{loc3}
most notably the absolute equilibrium ensemble (AEE) theory
\cite {Kraichnan67,Kraichnan75,FM76}, simply ignored all topological 
invariants except quadratic ones, such as enstrophy and helicity in
hydrodynamics or magnetic helicity, cross helicity, and square vector
potential (in two dimensions) in magnetohydrodynamic turbulence. These
integrals were honored the special attention in part because of their
ruggedness (survivability under the approximation of a finite number of
Fourier modes), but mostly because of the convenience to handle
quadratic integrals.

Despite the dissatisfaction with such a reasoning (cf.~\cite{CF87}),
the attempts to incorporate all topological invariants in Gibbs
statistics have been less frequent, the examples including
Vlasov-Poisson system \cite{Lynden-Bell67} and 2D Euler turbulence
\cite {Kuzmin82,Miller90,RS91,MWC92}.  For the reasons discussed in
Sec.~\ref{sec:conclusion} and Appendix~D, these attempts appear not
quite successful, because the non-Gaussianity of turbulent
fluctuations in these systems poses a fundamental difficulty in making
quantitative predictions about the equilibrium turbulent states.
\label{loc4}
The key problem is that the averaging with respect to the given
probability functional of a turbulence involves integration in a
functional (infinite dimensional) space.  These averages are well
defined---that is, independent of the discretization procedure
involved, if and only if the probability functional is Wienerian, or
Gaussian in the highest-order derivative \cite{Isichenko95}.
Otherwise, the result of the averaging is sensitive to the arbitrary
choice of the sequence of discrete representations, which makes such
probability functionals ambiguous, to say the least.
In the present work we point out that these difficulties are absent
from a certain class of systems; namely, those where all integrals of
motion are not higher than quadratic in the highest-order-derivative
variable.  Important examples of such systems, allowing a valid
Gibbs-ensemble description of turbulence, include two dimensional and
reduced magnetohydrodynamics.
 
The problem of accounting for all invariants is circumvented by the
representation of turbulence in the form of a gas of point vortices,
which is a very singular, and very special topologically, although an
asymptotically exact solution of the hydrodynamic equations.  The
localization of vorticity in point vortices makes the topological
constraints trivially fulfilled for any motions.  The conservation of
only energy and the Liouville theorem expressed in the convenient form
of the spatial variables of the vortices yield nicely to the
statistical mechanical description, although the thermodynamic limit
of infinitely many point vortices has long been a controversial issue
\cite {Onsager49,MJ73,ET74,FR82,BKH91}.  It remains unclear to what
extent the gas of many point vortices represents a continuum two
dimensional turbulence.  The frustrating dependence of the statistics
of point vortices on the arbitrary choice of their strengths was noted
by Onsager and reflects the above-mentioned fundamental difficulty in
the 2D Euler turbulence.
 
We pursue an analytical approach to continuum two dimensional ideal MHD
turbulence where all topological invariants are respected.  We use the
Gibbs ensemble analysis to predict the following evolution of the
turbulence.  An initial state evolves into (a)~a stationary, stable
coherent structure, which appears as the ``most probable state'' and
(b)~small-scale turbulence (fluctuations) with Gaussian statistics.  
\label{loc5}
The Gaussianity of the MHD fluctuations was recently numerically
confirmed by Biskamp and Bremer \cite {BB93}.  At large time the
fluctuations of the magnetic flux and the fluid stream function assume
vanishing amplitude and length scale (while containing finite energy
and dominating phase volume), and become essentially invisible on the
background of the coherent structure.  In this sense, the coherent
structure can be regarded as an attractor or a relaxed state, although
the underlying dynamics is perfectly Hamiltonian.
 
The concept of ``statistical attractor,'' introduced by Vladimir Yankov
\cite {KY80,PY89,DZPSY89}, emphasizes the method of analysis and describes
this kind of Hamiltonian relaxation, when the excess of phase-space
volume and energy get hidden in obscure (small-scale) corners of the
infinite dimensional phase space.  The fundamental difference between
the statistical attractors in nonintegrable wave-type systems
\cite {KY80,PY89,DZPSY89,Gruzinov93a} (where there are only a finite number
of integrals of motion) and the hydrodynamic-type systems (where the
number of integrals is infinite) is the universal shape of the
attractor---soliton---in the first case, and the nonuniversal shape of
the coherent structure---vortex---in the second case.  The appearance
of the coherent structure depends on the initial condition, but is the
same for all initial conditions with identical topological invariants
[Eqs.~({6}) and ({7})].  In this sense, the asymptotically emerging
coherent structures (relaxed states) in hydrodynamic systems are
attractors only within a certain subclass (basin) of initial
conditions.  These relaxed states can be called {\em topological
attractors}.
 
The specific of two dimensional MHD is that the coherent structure
inherits from the initial state all magnetic topology invariants, but
only a fraction of energy, the rest of which goes to the Gaussian
fluctuations.  These invariant sharing properties can be interpreted in
terms of the well-known reasoning of turbulent cascades. In the case of
2D MHD the energy cascade is direct (i.e., toward the small-scale
fluctuations), while the magnetic topology cascades inversely (toward a
large-scale magnetic structure). However, compared to the cascade
description, the invariant sharing properties appear to present a
clearer physics of what happens to the conserved quantities in a closed
turbulent system.  In fact, this allows us to predict the appearance of
the relaxed state, which should minimize energy subject to certain
topological constraints.  One of the novelties of our analysis is using
a second functional set of ``cross'' topology invariants \cite
{MH84,IM87}, which was not used in MHD turbulence theories so far,
including the previous note \cite {Gruzinov93b}.  We find that these
invariants have an important effect on the statistical description of
turbulence; specifically, the shape of the coherent structure is
sensitive to the cross topology invariants. In many features our
development is analogous to the Gibbs ensemble treatment of the
truncated Fourier representation of 2D MHD which was investigated by
Fyfe and Montgomery \cite{FM76}. The reason for this resemblance is
partial consistency of the Gibbs statistics with partial invariants
accounting. In fact, our theory shows that truncated 2D MHD equations
partially represent the true statistics of ideal 2D MHD.
 
Unlike neutral fluid dynamics, magnetohydrodynamics in two dimensions
are known to produce energetic small scales.  This makes the difference
between 2D and 3D for MHD turbulence less drastic than that for fluid
turbulence.  In our model, we observe two types of small-scale
behavior: (a)~for a generic (that is, topologically nontrivial) initial
condition, the coherent structure must have discontinuities in the form
of current sheets and (b)~the Gaussian fluctuations in the long evolved
state have both vanishing length scale and amplitude so that the
gradients and the energy are finite.  The numerically observed
current-sheet-type structures \cite{BW89} are explained by our theory
as the singular coherent structures.
 
The paper is organized as follows.  In Sec.~\ref{sec:equations} we
present the equations and the constants of the motion.  In
Sec.~\ref{sec:stationary} we discuss the stationary solutions of the
MHD equations and formulate the Arnold variational principle in the
form suitable for 2D MHD.  In Sec.~\ref{sec:gibbs} the canonical
ensemble approach to MHD turbulence is set forth
(Sec.~\ref{sec:ensemble}), and the most probable state
(Sec.~\ref{sec:coherent}) and the fluctuations about this state
(Sec.~\ref{sec:fluctuations}) are analyzed.  
The key issue of how the integrals of motion are divided between the
coherent structure and the fluctuations is addressed in
Sec.~\ref{sec:partition}.  In Sec.~\ref{sec:relaxation} we reformulate
the properties of the coherent structure using a variational principle
of iso-topological relaxation, which allows us to predict the
appearance of the structure.  This prediction is then compared with
numerical results (Sec.~\ref{sec:numerical}).  In
Sec.~\ref{sec:dissipation} we speculate on the role of small
dissipation and the relation between the resistive and the ideal MHD
relaxation in the kink tearing mode in tokamaks.
Section~\ref{sec:conclusion} restates the principal steps of our
statistical method and summarizes our work.  Some technical details and
results not directly related to MHD turbulence are set in Appendices.
In Appendix~A we discuss the Lyapunov stability of MHD and Euler fluid
equilibria and point out the relation of minimum- and maximum-energy
stability to positive- and negative-temperature Gibbs states,
respectively.  Appendix B addresses the Liouvillianity of the
eigenmodes that we use in the Gibbs statistics.  In Appendix~C the
spectrum of the eigenmodes is studied.  In Appendix~D we discuss the
application of the Gibbs-ensemble formalism to the turbulence of two
dimensional Euler fluid.

\section{Equations and constants of the motion}
\label{sec:equations}
 
We consider the set of equations of two dimensional incompressible 
ideal magnetohydrodynamics (cf.~\cite {TMM86})
 \bqy
\p_t a &=& \{\ps,a\}\ ,
\label{1} \\
\p_t\om &=& \{\ps,\om\} + \{j,a\}\ ,
\label{2}
\eqy 
where $\{A,B\}\equiv\bfna A\times\bfna B\cdot\wh\bfz$ denotes
the Poisson bracket, $\wh\bfz$ the unit vector in the $z$ direction,
$\ps(x,y,t)$ the stream function of the fluid velocity field
$\bfv=\bfna\times(\ps\wh\bfz),\; \om=-\bfna^2\ps$ the fluid vorticity,
$a(x,y,t)$ the normalized vector potential of the magnetic field
$\bfB=(4\pi\rh)^{1/2}\bfna\times(a\wh\bfz)$ ($\rh$ being the constant fluid
density), and $j=-\bfna^2a$ the normalized electric current flowing
perpendicular to the $(x,y)$ plane.  The boundary conditions $a=\ps=0$ are
assumed at a rigid boundary encompassing the finite domain
with the area $\cals$.
 
The incompressibility of the fluid motion is a reasonable approximation
in tokamaks where the strong toroidal magnetic field $B_z$ makes plasma
compression energetically expensive.  In the reduced MHD approximation
\cite {KP73,RMSW76,Strauss76}, where the fast \Alfven\ and magnetosonic
waves due to $B_z$ are ignored, the strong uniform field $B_z$ drops
out of the equations of the motion.
 
The system ({1})---({2}) conserves the following quantities:  the energy
 \bq 
E = E_m + E_f =
\frac{1}{2}\int\left[ (\bfna a)^2 + (\bfna\ps)^2 \right]\,d^2\bfx =
\frac{1}{2}\int\left( aj + \ps\om \right)\,d^2\bfx\ ,
\label{3}
\eq 
consisting of the magnetic part $E_m$ and the fluid part $E_f$, 
the momentum (with translationally invariant, or in the absence of, 
boundaries)
 \bq
\bfP = \int\bfv\,d^2\bfx=
\frac{1}{2}\int\bfx\times\wh\bfz\om\,d^2\bfx\ ,
\label{4}
\eq 
the angular momentum (with circular or no boundaries)
 \bq
M\wh\bfz = \int\bfx\times\bfv\,d^2\bfx=
\frac{1}{3}\int\bfx\times(\bfx\times\wh\bfz\om)\,d^2\bfx\ ,
\label{5}
\eq
the magnetic topology invariants
 \bq
I_F = \int F(a)\,d^2\bfx\ ,
\label{6}
\eq
and the ``cross'' topology invariants
 \bq
J_G = \int \om G(a)\,d^2\bfx = \int G'(a)\bfna\ps\cdot\bfna a\,d^2\bfx\ ,
\label{7}
\eq
where $F$ and $G$ are arbitrary functions.  Along with the continuum
set of integrals ({6}) and ({7}), we will also use their discretized
analogues,
 \renewcommand{\theequation}{\arabic{equation}a} 
 
 \addtocounter{equation}{-2}
 \bqy
I_n &=& \int a^n\,d^2\bfx\ ,
\label{6a}
\\
J_n &=& \int \om a^n\,d^2\bfx\ ,
\label{7a}
\eqy
\renewcommand{\theequation}{\arabic{equation}}
through which the continuum invariants can be Taylor expanded.
 
Strictly speaking, invariants ({6}) and ({7}) do not yet imply the
conservation of topology.  Equation ({6}) only means that allowed
motions are incompressible interchanges of fluid elements together with
their ``frozen'' values of the magnetic flux $a$.  The topology of the
contours of $a$ will be conserved only if these interchanges are
performed by continuous movements, a constraint which is not built into
Eq.~({6}) but follows from the equations of motion for smooth initial
conditions.  Then the conservation of magnetic topology expressed by
Eq.~({6}) means that if the contour $a(x,y)=a_1$ initially lies inside
the contour $a(x,y)=a_2$, then this topological relation is preserved
by the motion; if a contour of $a$ has a hyperbolic (saddle, $x$)
point, this quality will also persist.  In addition to the topological
constraints, the integrals ({6}) also specify the incompressibility of
the fluid, so that the area inside a given contour of $a$ remains
constant.  The geometrical meaning of the integrals ({7}) is that the
amount of the fluid vorticity $\om$ on a given contour of $a$ (to be
more precise, the integral of $\om$ over the anulus between two
infinitesimally close contours) is conserved.  Although stating
nothing of the contours of $\om$ or $\ps$, the conservation of the
integrals ({7}) also bears certain topological relation between the
magnetic field and the vorticity field, which motivates our notation of
the ``cross topology invariants.''
 
The invariants $J_G$ appear to be poorly known, although the particular
member of the family ({7a})---the cross helicity
 \bq
J_1 = \int \om a\,d^2\bfx = 
\frac{1}{\sqrt{4\pi\rh}}\int \bfv\cdot\bfB\,d^2\bfx
\label{8}
\eq
---has been extensively discussed in the literature.  The invariants
({7}) were first noted by Morrison and Hazeltine \cite {MH84}.
Independently, a similar set of integrals was used to study the vortex
stability in the framework of two dimensional electron MHD \cite
{IM87}.  The idea towards the existence of a second set of topological
invariants is suggested by the observation that there is another
frozen-in quantity, namely the vorticity $\om$, in the Euler limit
$a\equiv0$, which must have a counterpart in the MHD case $a\ne0$.
Once the existence of a second functional set of invariants is
suspected, it is not hard to guess the form of the topological
invariants ({7}).  Although this is not straightforward, one can trace
the transition, as $a\to0$, from Eqs.~({{6}) and ({7}) to the Euler
invariants $\int F(\om)\,d^2\bfx$.
 
Another way to find the topological invariants is to identify the
Hamiltonian structure using a noncanonical Poisson bracket \cite
{MH84}, whereby the topological invariants appear as Casimirs.

\section{MHD equilibria and Arnold's variational principle}
\label{sec:stationary}
 
The system of equations ({1})---({2}) has stationary solutions 
satisfying
 \bqy
\{\ps,a\} &=& 0\ ,
\label{9}
\\
\{\ps,\bfna^2\ps\} &=& \{a,\bfna^2a\}\ .
\label{10}
\eqy
The equilibrium condition can be rewritten in a more convenient form by
substituting the functional dependence $\ps=\ps(a)$, which is implied by
Eq.~({9}), into Eq.~({10}).  Then, after simple manipulations, we find
 \bq
\{\Ps(a),\bfna^2\Ps(a)\}=0\ ,
\label{9a}
\eq
where
 \bq
\Ps(a)=\int_0^a da'\sqrt{\pm\left(1-[d\ps(a')/da']^2\right)\,}\ ,
\label{9b}
\eq
and the sign is chosen to make the square root real.  As equation
(11) shows, any two dimensional MHD equilibrium with fluid flow
($\ps\ne0$) is reduced to a purely magnetic ($\ps=0$) equilibrium for a
modified magnetic vector potential $a'=\Ps(a)$.  Note that the magnetic
field lines (the contour lines of the vector potential) are identical
for both the true field $a$ and the modified field $a'$, although the
values of $a$ and $a'$ on these lines are different.
 
The most evident stationary solution is given by an arbitrary circular
distribution $\ps=\ps(r),\;a=a(r)$, where $r$ is the distance from the
origin.  Another solution to ({9})---({10}) corresponds to the
identical zero in one of the Els\"asser variables, where
$\ps(x,y)\equiv\pm a(x,y)$ is an arbitrary function of $x$ and $y$, so
that $\Ps(a)\equiv0$.  For the case when $|\ps|$ and $|a|$ are not
identical, there exist many periodic and quasiperiodic solutions in the
form
 \bq
C\ps(\bfx)=a(\bfx)=\sum_{m=1}^{N}A_m\cos(\bfk_m\bfx+\theta_m)\ ,
\label{11}
\eq
where the moduli of the wavevectors $\bfk_m$ are the same.  In addition
to the smooth solutions one can devise a wide class of singular
solutions with appropriate boundary conditions at the lines of
discontinuity.  Without additional physical constraints, such as
stability or topology, the class of all equilibria is too wide to be
useful.  In Secs.~\ref{sec:gibbs} and \ref{sec:relaxation} we provide
such constraints, which specify the physically interesting (attracting)
equilibria.  In many cases these equilibria must be singular.
 
There exists a profound relation between the stationary solutions and
the constants of the motion.  For finite dimensional conservative
systems, the D'Alembert variational principle says that the energy
variation be zero at an equilibrium.  For a hydrodynamic-type
conservative system, the counterpart of the D'Alembert theorem is the
Arnold variational principle.  Originally formulated for the Euler
equation, but carried over without difficulty to other hydrodynamics,
the principle states that at a stationary (and only stationary)
solution the variation of energy, {\em subject to the conservation of
all topological invariants}, be zero:
 \bq
\de E\bigg|_{I_F,J_G=\const,\ \forall F {\ \rm and\ } G}=0\ .
\label{arnold}
\eq
If, in addition, the second variation is definite---that is, the energy
assumes a nondegenerate conditional extremum, then the equilibrium is
Lyapunov stable.  In fact, this was the search for stable fluid flows
which motivated Arnold's work.  The power of the method lies in the
possibility to write the general iso-topological (iso-vortical in
Arnold's notation) variation in a closed form. When all the integrals
({6}) and ({7}) are to be conserved, such a variation is 
 \bq
\De\left(\begin{array}{l}  a\\
                           \om 
         \end{array}\right) =
\de\left(\begin{array}{l}  a\\
                           \om 
         \end{array}\right) + 
\frac{1}{2!}
\de^2\left(\begin{array}{l} a\\
                            \om 
           \end{array}\right) + \ldots\ ,
\label{12}
\eq
where the infinitesimal iso-topological variation is given by
\cite {IM87}
 \bq
\de\left(\begin{array}{c} a\\
                          \om 
         \end{array}\right) =
\left(\begin{array}{c} \{\mu,a\}\\
                       \{\mu,\om\} + \{\nu,a\} 
      \end{array}\right)\ ,
\label{13}
\eq
$\mu(x,y)$ and $\nu(x,y)$ being arbitrary functions.  The operator of
finite variation $\De$ can be symbolically expressed through the
infinitesimal variation $\de$ as 
 \bq
\De = \exp\de - 1\ .
\label{14}
\eq
It is easy to verify that the finite variation (15) conserves both sets
of integrals $I_F$ and $J_G$ to all orders in $\mu$ and $\nu$.  The
form of the variation is suggested by the form of Eqs.~({1})---({2}),
where one can substitute the quantities $\ps$ and $j$ by arbitrary
$\p\mu/\p t$ and $\p\nu/\p t$, respectively, to preserve only the
Poisson-bracket structure of the equations and thereby to
iso-topologically (that is, at constant $I_F$ and $J_G$) drag the
fields $a$ and $\om$ to a new state, where the values of all other
integrals, if any, are generally different from those of the initial
state.
 
Let us see what happens to the energy ({3}) under the variation
(15)---(16).  Writing the total change in the energy in the form 
$\de E+\de^2E/2!+\ldots$, we obtain after integrating by parts
 \bq
\de E = -\int[\mu\left(\{\ps,\om\}+\{j,a\}\right)+ 
\nu\{\ps,a\}]\,d^2\bfx\ .
\label{15}
\eq
By requiring that the first energy variation ({18}) be zero for all
$\mu$ and $\nu$ we arrive exactly at the system of equations
({9})---({10}) specifying the equilibrium solution.  This strongly
suggests that the two sets of the topological invariants ({6}) and
({7}) are indeed complete, a condition necessary to apply statistics
(Sec.~\ref{sec:gibbs}) in a meaningful way.  
 
The second iso-topological variation of energy can be used to
investigate the stability of MHD equilibria, as discussed in
Appendix~A.
 
\section{Gibbs statistics of two dimensional MHD turbulence}
\label{sec:gibbs}
 
We now wish to analyze the long-time evolution of the system subject to
Eqs.~({1}) and ({2}) for a given initial condition
$a_0(\bfx),\;\ps_0(\bfx)$.  The probability distribution functional
$P[a(\bfx),\ps(\bfx)]$ can serve this purpose.  This functional
specifies the relative probability, with respect to time measure, of
the spatial behaviors of various states $a(\bfx,t)$ and $\ps(\bfx,t)$.
As $P$ is invariant under the evolutional change of the fields $a$ and
$\ps$, it must be a function of the constants of motion ({3})---({7}).

\subsection{Choice of statistical ensemble}
\label{sec:ensemble}
 
For a conservative Hamiltonian system, $P$ is given by the microcanonical
ensemble,
 \bq
P_{MC}[a,\ps]=
\de(E[a,\ps]-E_0)\,\prod_n\de(I_n[a,\ps]-I_{n0})\,\de(J_n[a,\ps]-J_{n0})\ ,
\label{MC}
\eq
specifying a uniform distribution on the manifold of the specified
(initial) integrals of motion.  It must be emphasized that the validity
of the microcanonical ensemble requires at least four assumptions.  
 \begin{enumerate}
 \item 
The phase space must be finite dimensional; that is, the fields $a$ and
$\ps$ are parameterized by discrete dynamical variables 
$f_m,\ m=1,2,\ldots,N$, so that the concept of measure in the space of
states is meaningful.  This is a tricky issue as to how many variables
are needed (see discussion in Sec.~\ref{sec:introduction}). 
 \item 
The motion on the manifold of conserved invariants is nonintegrable
(chaotic).  The fundamental phenomenon lying behind the ergodic
behavior (uniform distribution over the manifold) is Hamiltonian chaos,
whose principal manifestation is the exponential divergence of nearby
trajectories.  The chaotic motion is possible only if the
dimension of the manifold ($N-N_I$, where $N_I$ is the number of
invariants) is three or more (four or more to allow the Arnold
diffusion, so that all of the manifold might be visited by each
trajectory).  The property of ergodicity was proved for very special
cases \cite {Sinai70} but is believed to be generally valid if the
dimension of the manifold is sufficiently large: $N-N_I\gg1$.  The
remarkable accuracy of the classical thermodynamics is connected with
the macroscopic numbers of degrees of freedom ($N\sim10^{23}$) and only
a few invariants.  It is also natural to expect that the microcanonical
statistics will work in turbulence (which can be loosely defined as
chaos in PDE), where the limit $N\to\infty$ should be carefully taken.
 \item 
The manifold of conserved integrals specified by the delta functions in
({19}) must be connected.  The problem of connectivity of complicated
iso-surfaces is related to the percolation problem \cite {Isichenko92}.
The percolation threshold, above which the connectivity takes place, is
inversely proportional to the dimension of the iso-surface.  This
suggests that in the continuum (infinite dimensional) limit the
connectivity of the manifold of conserved integrals should not be a
problem.
 \item 
The dynamical variables $f_m(t)$ must satisfy the Liouville theorem:
$\sum_m\p\dot f_m/\p f_m=0$.  For non-Liouvillian variables a weighting
factor (the Jacobian of change to Liouvillian variables) should be
included in Eq.~({19}).  
 \end{enumerate}
 
In a hydrodynamic-type system, where the number of dynamical
constraints is infinite, we encounter another difficulty.  Namely, any
attempt to restrict the dimension of the phase space without
restriction on the number of conserved quantities immediately drives
the manifold of conserved integrals into an empty set, where no mixing
may occur.  Motivated by the experimental/numerical observation that
turbulence does exist, as well as by the functional arbitrariness of
the iso-topological variation ({16}), we adopt a hypothesis that there
exists a ``meaningful'' $N$ dimensional MHD approximation with at most
$N_I(N)$ conserved integrals, where both $N_I$ and $N-N_I$ go to
infinity as $N\to\infty$.  (In Zeitlin's example \cite {Zeitlin91} for
the Euler fluid $N_I\simeq N^{1/2}$.) The specific form of this
approximation is unimportant for our arguments.
 
The microcanonical ensemble ({19}) is inconvenient to handle and is
commonly transformed into the more convenient canonical (Gibbs)
ensemble by integrating $P_{MC}$ over most of the dynamical variables
in the amount of $N_{th}\gg N-N_{th}\gg1$.  These $N_{th}$ degrees of
freedom can referred to as ``thermal bath.'' The integration over the
thermal bath variables leads to an exponential dependence of the
resulting distribution on the integrals of motion expressed through the
remaining $N-N_{th}$ variables, the rest of the information being
stored in arbitrary constants called temperatures.
 
In our problem, the dimension $N'=N-N_{th}$ of the subsystem can
be also taken large, which amounts to another finite dimensional ($N'$)
MHD approximation.  However, now we have the canonical distribution
over the remaining $N'$ variables,
 \bq
P[a,\ps] = 
\exp\left[-\al\left(
E[a,\ps] + \sum_n\be_n I_n[a] + \sum_n\ga_n J_n[a,\ps]
\right)\right]\ ,
\label{canonical}
\eq
instead of the microcanonical one ({19}).  A drastic simplification
achieved by the change of the ensemble is that in the
finite dimensional Gibbs distribution ({20}), where the fields $a$ and
$\ps$ are parameterized by $N'$ modes and the summation over invariants
runs up to $N_I(N')$, we may extend the summation up to infinity
without significant change in the result, which was impossible for the
product of delta functions in the microcanonical ensemble ({19}).
 
In Eq.~({20}), the constants $\al,\;\al\be_n$ and $\al\ga_n$ appear
as the reciprocal temperatures corresponding to each invariant.  These
constants are to be determined from the initial state by solving
the infinite system of equations:
 \bq
\left<E\right>_P=E_0\ ,
\quad
\left<I_n\right>_P=I_{n0}\ ,
\quad
\left<J_n\right>_P=J_{n0}\ ,
\quad n=0,1,2,\ldots\ , 
\label{conservation}
\eq
expressing the conservation of the integrals of motion.  Here the
subscript ``0'' refers to the initial state.  As a result of solving
Eqs.~({21}), each parameter $\al,\;\be_n,$ or $\ga_n$
($n=0,1,2,\ldots$) is a function of the infinity of the initial
invariants $E_0,\;I_{n0}$ and $J_{n0}$.  It is emphasized that there is
no arbitrariness in the temperatures characterizing the Gibbs
distribution of a closed system.  In fact, this is the central problem
in our theory how to determine these temperatures in order to predict
the final state from the given initial state.
 
The angular brackets in Eqs.~({21}) denote the ensemble averaging,
 \bq
\left<A\right>_P=
\frac{\int A\, P[a,\ps]\cald a\cald\ps}{\int P[a,\ps]\cald a\cald\ps}\ , 
\label{averaging}
\eq
which involves functional integrals over the space of the system
states.  This kind of integrals do not always exist.  
\label{loc6}
However, when the probability functional $P$ is Gaussian, the
functional integrals belong to the important class of Wiener integrals
\cite {Wiener58} (their complex counterparts are known as path integrals
\cite {FH65}), which are soluble and well-behaved.  This is exactly
what we use in order to resolve the ultraviolet catastrophe.  In fact,
we seek the long evolved state in the form of a coherent structure plus
small-amplitude fluctuations.  This allows us to expand the integrals
in the exponential ({20}) about the coherent structure up to quadratic
terms, which will result in a Gaussian probability functional.
 
The uniform and additive (with respect to the eigenmodes) invariants is
another assumption lying behind the transition from the microcanonical
({19}) to the canonical ({20}) distribution functionals.  The
additivity of the invariants can be achieved by the procedure of
diagonalization, which only works for quadratic forms.
 
In the spirit of the conventional statistical mechanics we call the
state specified by the detailed list of variables $(f_1,\ldots,f_N)$
{\em the microstate}, whereas the union of all microstates with the
same $(f_1,\ldots,f_{N'}),\ N'\ll N$ {\em the macrostate}.  The entropy
$S$---a functional of the macrostate---is then introduced as the logarithm
of the number of various microstates corresponding to the given
macrostate.  Up to an additive constant, $S$ is the logarithm of the
microstate phase volume on the manifold ({19}): 
 \bq
S[f_1,\ldots,f_{N'}]=
\ln\int P_{MC}[f_1,\ldots,f_N]\,df_{N'+1}\ldots df_N\ .
\label{S1}
\eq
In other words, the entropy is simply the logarithm of the canonical
distribution functional ({20}),
 \bq
S[a,\ps]=\ln P[a,\ps] = -\al(E[a,\ps]+I_\be[a]+J_\ga[a,\ps])\ ,
\label{S2}
\eq
if $N'\gg1$.  In Eq.~({24}), the integrals $I_\be$ and $J_\ga$ are defined
by Eqs.~({6}) and ({7}), respectively, and the functions 
 \bq
\be(a)=\sum_n\be_na^n \quad{\rm and}\quad
\ga(a)=\sum_n\ga_na^n
\label{be_and_ga}
\eq
can be regarded as ``topological temperature functions.''  
 
It is emphasized that the macrostate specified by $N'$ degrees of
freedom can be arbitrarily detailed, as we may let $N'\to\infty$ (while
preserving the requirement $N'\ll N$), so that formally there is little
difference between macrostates and microstates in a continuum system,
although the apparatus of the canonical distribution ({20}) and the
entropy ({24}) is much more convenient than that of the microcanonical
ensemble ({19}).

\subsection{Coherent structure: the most probable state}
\label{sec:coherent}
 
Maximizing the probability ({20}) or, equivalently, the entropy
({24}) yields ``the most probable state'' of the system. 
Upon varying $S$ with no restriction on the field variations $\de a$ and
$\de\ps$ we obtain
 \bq
\de(E[a,\ps]+I_\be[a,\ps]+J_\ga[a,\ps])=0\ ,
\label{MPS}
\eq
which results in a stationary solution $\left(a_s(\bfx),\ps_s(\bfx)\right)$
satisfying
 \bqy
\bfna^2a_s &=& \be'(a_s) + \ga'(a_s)\bfna^2\ga(a_s)\ ,
\label{25}
\\
\ps_s &=& -\ga(a_s)
\label{26}
\eqy
[compare with ({9})---({10})].  There is nothing surprising in that the
most probable state is stationary, because varying a linear combination
of the energy and the topological integrals ({26}) amounts to the
Arnold (iso-topological) variation written with the help of Lagrange
multipliers.  The relation (28) stating that the fluid flow is along
the magnetic field lines is characteristic of the ``dynamic alignment''
developing in course of turbulent MHD relaxation \cite {TMM86}.
 
Similarly to the transformation (11)---(12), we can rewrite
Eqs.~(27) and (28) in the form
 \bq
\bfna^2\Ga(a_s)=\be'(a_s)/\Ga'(a_s)\ ,
\label{26a}
\eq
where
 \bq
\Ga(a_s)=\int_0^{a_s} da\sqrt{1-[\ga'(a)]^2\,}\ .
\label{26b}
\eq
This representation of the coherent structure will be used in
Sec.~\ref{sec:relaxation}.
 
The quantity 
 \bq
\ga'(a_s(\bfx))=-\frac{\bfna\ps_s}{\bfna a_s}=
-\frac{\bfv_s}{\bfB_s/\sqrt{4\pi\rh}} 
\label{Mach}
\eq
is the local Mach number of the fluid flow.  Although some interesting
phenomena may occur near the lines where $|\ga'|=1$, we will restrict
our attention to the sub-\Alfvenic\ case $|\ga'|<1$.  A sound
motivation for this is found in the absence of maximum-energy states in
2D magnetohydrodynamics (see Appendix~A) and the relaxation of
turbulence to minimum-energy states where $|\ga'|$ is necessarily less
than one [see Eq.~({44}) and Appendix~C].  Even though the initial
condition is highly super-\Alfvenic, $|\bfna\ps_0|\gg|\bfna a_0|$, the
necessary magnetic field will be generated by means of turbulent
dynamo.
 
In addition to equations (27) and (28), we must require that the
equilibrium state $(a_s,\ps_s)$ be actually the maximum of the entropy
$S$.  This requirement, which is pursued in the next subsection, means
that the coherent structure must be Lyapunov stable, which is natural
to expect of a relaxed state.  Indeed, the ``fine-grained entropy''
$S$, as defined by Eq.~({24}) is an integral of motion playing the role
of a Lyapunov functional.

\subsection{Fluctuations:  the Gaussian turbulence}
\label{sec:fluctuations}
 
Now that we have identified (or, rather, assumed the presence of) the
coherent structure $(a_s,\ps_s)$, we seek solution to the problem
({1})---({2}) in the form
 \bq
a(\bfx,t)=a_s(\bfx)+\wt a(\bfx,t)\ ,\quad 
\ps(\bfx,t)=\ps_s(\bfx)+\wt\ps(\bfx,t)\ ,
\label{decomposition}
\eq
where the amplitude of the fluctuations $\bff=(\wt a,\wt\ps)$ is
expected (and below confirmed) to be small in the long-time limit.
With this in mind, we calculate the second variation of the entropy:
 \bq
\de^2S = -\al\int\Big[
(\bfna\wt a)^2 + (\bfna\wt\ps)^2 + \be^*\wt a^2 + 2\ga'(a_s)\wt a\wt\om
\Big]\,d^2\bfx\ ,
\label{27}
\eq
where we denote $\de a=\wt a,\;\de\ps=\wt\ps$, and
$\be^*\equiv\be''(a_s(\bfx))+\om_s(\bfx)\ga''(a_s(\bfx))$.  In order
for the fluctuations to be finite, $\de^2S$ must be negative definite.
The integral quadratic form on the right hand side (RHS) of Eq.~(33)
can be represented as the matrix element $\left<\bff|W|\bff\right>$ of
the linear self-adjoint tensor operator,
 \bq
W\left(\begin{array}{l}
         \wt a \\
         \wt\ps
       \end{array}\right)=
\left(\begin{array}{rr}
         (\be^*-\bfna^2)         & \quad-\ga'\bfna^2 \\
         -\bfna^2(\ga'\ldots)    & \quad-\bfna^2
\end{array}\right)
\left(\begin{array}{l}
         \wt a \\
         \wt\ps
       \end{array}\right)
\equiv
\left(\begin{array}{c}
         \be^*\wt a+\wt j+\ga'\wt\om \\
         -\bfna^2(\ga'\wt a) + \wt\om
\end{array}\right)\ ,                          
\label{W}
\eq
acting on a pair of functions $\bff=(\wt a,\wt\ps)$.  The boundary
conditions are $\wt a=0$ (tangential magnetic field) and $\wt\ps=0$
(tangential velocity) at the boundary of the finite domain.
 
The orthonormal set of the eigenfunctions $(a_m,\ps_m)$ of $W$ provides a 
natural representation of the fluctuations:
 \bq
W\left(\begin{array}{l}
         a_m\\
         \ps_m
       \end{array}\right)=\la_m
\left(\begin{array}{l}
         a_m\\
         \ps_m
       \end{array}\right)\ ,                          
\label{eigen}
\eq
The standard definition of the orthonormality implies
 \bq
\int(a_ma_n+\ps_m\ps_n)d^2\bfx=\de_{mn}\ .
\label{orthonormality}
\eq
Upon expanding the fluctuation field 
 \bq
\bff(\bfx,t)=\sum_m
f_m(t)\left(\begin{array}{l}
              a_m(\bfx) \\
              \ps_m(\bfx)
            \end{array}\right)\ ,                          
\label{28b}
\eq
in a series over the complete set of the eigenfunctions, the probability
distribution of the fluctuations is conveniently written as
 \bq
P[\bff]\equiv\exp\left(\frac{\de^2S}{2}\right) = 
\exp\left(-\frac{\al}{2}\sum_m\la_m f_m^2\right)\ ,
\label{Gaussian}
\eq
which is a Gaussian distribution.  
 
In Appendix B we discuss the Liouvillianity of the variables $f_m$ and
show that the averages over the distribution ({38}) are done by
replacing $\cald a\cald\ps$ by $\prod_mdf_m$ in Eq.~({22}).  Then the
fundamental averages are
 \bq
\left<f_m\right>=0\ ,\quad \left<f_mf_n\right>=\de_{mn}/(\al\la_m)\ .
\label{averages}
\eq
 
The eigenvalues $\la_m$ depend on the temperature parameters
$\al,\;\be_n$, and $\ga_n$ and, through those, on the initial state.
However, the behavior of the eigenvalues becomes universal in the
ultraviolet ($m\gg1$) limit.  In Appendix~C we show that the spectrum
of the matrix operator $W$ is similar to that of the standard scalar
Schr\"odinger operator $U(\bfx)-\bfna^2$, whose quasiclassical
eigenvalues are determined by the Bohr-Sommerfeld quantization rule:
 \bq
\la_m^\pm\simeq C^\pm\frac{4\pi m}{\cals}\ ,\quad m\gg k_s^2\cals\ ,
\label{spectrum}
\eq
where $\cals$ is the area of the domain and $k_s$ the characteristic
wavenumber of the smooth part of the coherent structure.  (In
Sec.~\ref{sec:relaxation} we show that the coherent structure can also
have singularities---current sheets---which are not important in this
context.) In the Schr\"odinger case the constant $C$ in Eq.~({40}) is
unity, whereas for the operator ({34}) there are two branches of
eigenmodes with
 \bq
1-|\ga'|_\smax<C^-<C^+<1+|\ga'|_\smax\ .
\label{Cpm}
\eq
As also shown in Appendix~C, the eigenfunctions behave in the
ultraviolet [Wentzel-Kramers-Brillouin (WKB)] limit as
 \bq
\ps_m^\pm(\bfx)\simeq\pm a_m^\pm(\bfx)\ ,\quad m\gg k_s^2\cals\ ,
\label{eigenfunctions}
\eq
and the WKB wavenumber of the $m$th mode is
 \bq
\bfk_m^2\simeq4\pi m/\cals\ ,\quad m\gg k_s^2\cals\ .
\label{wavenumber}
\eq
 
The maximum of entropy ($\de^2S<0$) is equivalent to the
non-negativeness of all eigenvalues $\la_m$ of operator ({34}).  It
appears difficult to formulate the exact criterion of the positive
definiteness of $W$ in the general case; however, a sufficient
condition can be derived by applying the Silvester criterion to the
integrand in (33) considered as a plain quadratic form of fifth order
expressed through the variables $\wt a,\ \bfna\wt a$, and
$\bfna\wt\ps$.  Then the result is
 \bq
\al>0\ ,
\quad
|\ga'|<1\ ,
\quad
\be^*(1-\ga'^2)>(\bfna\ga')^2\ .
\label{44}
\eq
Under these (or perhaps milder) constraints, the coherent structure is
Lyapunov stable as realizing minimum of the conserved quantity
$-S/\al=E+I_\be+J_\ga$.

\subsection{Partition of conserved quantities between the coherent 
structure and the fluctuations}
\label{sec:partition}
 
So the long evolved state of the 2D MHD turbulence involves two constituents,
namely the stationary, stable coherent structure $(a_s(\bfx),\ps_s(\bfx))$ and
the fluctuations $(\wt a(\bfx,t),\wt\ps(\bfx,t))$ distributed according to the
Gaussian law ({38}).  The initial state's invariants are shared
between the structure and the fluctuations,
 \bq
E_0=E_s+\wt E\ ,\quad I_{F0}=I_{Fs}+\wt I_F\ ,\quad J_{G0}=J_{Gs}+\wt J_G\ ,
\label{sharing}
\eq
where the subscripts $0$ and $s$ refer to the initial state and the coherent
structure, respectively, and tilde to the fluctuations.  The Gaussianity and
the integral sharing properties ({45}) follow from our assumption of 
the small amplitude of the fluctuations, which we confirm below.  In addition, 
we establish that  $\wt I_F=0$, which bears a useful topological corollary.
 
We start with the fluctuation energy
 \bq
\wt E=\left<\frac{1}{2}
\int(\wt j\wt a+\wt\ps\wt\om) d^2\bfx\right>=
\frac{1}{2}
\sum_{mn}\left<f_mf_n\right>\int(j_ma_n+\ps_m\om_n) d^2\bfx\ .
\label{E1}
\eq
Using formula ({39}) and eliminating $j_m$ and $\om_m$ with the help
of Eqs.~({34}) and ({35}), Eq.~({46}) can be rewritten in terms of
only $a_m$ and $\ps_m$.  At $m\gg k_s^2\cals$ the principal term in the
fluctuation energy is
 \bq
\wt E=\frac{1}{2\al}\sum_m
\int\frac{a_m^2-2\ga'a_m\ps_m+\ps_m^2}{1-\ga'^2}d^2\bfx\ .
\label{E2}
\eq
The integrand in (47) is greater than $(a_m^2+\ps_m^2)/(1+|\ga'|)$ and
less than $(a_m^2+\ps_m^2)/(1-|\ga'|)$.  As the orthonormality condition
({36}) then implies, the sum ({47}) diverges with the number
of eigenmodes $N\gg1$ {\em linearly:}
 \bq
\wt E=\frac{C_NN}{2\al}\ ,\quad N\gg1, \quad
\frac{1}{1+|\ga'|_\smax}\le C_N\le\frac{1}{1-|\ga'|_\smax}\ .
\label{E3}
\eq
 
Equation ({48}) is a remnant of the equipartition of energy between the
degrees of freedom (eigenmodes).  At finite temperature the energy would
diverge as $N\to\infty$, which constitutes the well-known ``ultraviolet
catastrophe.''  However, $\wt E$ is bounded from above by the initial energy
$E_0$.  Therefore the energy temperature $1/\al$ of the fluctuations should
decrease with the number $N$ of the effectively excited modes.  From the
conservation of energy [Eq.~({45})] we infer
 \bq
\al=\frac{C_NN}{2(E_0-E_s)}\ ,\quad N\gg k_s^2\cals\ .
\label{al}
\eq
 
Analogously to the fluctuation energy, $\wt I_F$ and $\wt J_G$ also diverge at
constant $\al$, as $N\to\infty$.  However, the divergence of $\wt I_F$ is
only {\em logarithmic} in $N$, because the role of small scales is less
pronounced in the magnetic topology invariants ({6}), which involve no
derivatives of $a$ and $\ps$.  [The linear divergence of $\wt E$ and $\wt J_G$
is due to the terms $(\bfna a)^2$ and $\om=-\bfna^2\ps$ in ({3}) and
({7}), respectively.]  Similarly to ({46})---({48}), we have  
 \bq
\wt I_F=\frac{1}{2}\int F''(a_s)\left<\wt a^2\right>d^2\bfx=
\sum_m\frac{1}{2\al\la_m}\int F''(a_s)a_m^2\,d^2\bfx\ ,
\label{I1}
\eq
and, in accordance with Eqs.~({36}) and ({40}), 
 \bq
|\wt I_F|\le
\frac{|F''|_\smax}{2\al}\sum_m\frac{1}{\la_m}\le
\frac{|F''|_\smax\cals}{8\pi\al(1-|\ga'|_\smax)}\,\ln N\ .
\label{I2}
\eq
Upon substituting expression ({49}) into Eq.~({51}) we
obtain
 \bq
|\wt I_F|\le
\frac{|F''|_\smax E_0\cals}{4\pi}\,
\frac{1+|\ga'|_\smax}{1-|\ga'|_\smax}\,
\frac{\ln N}{N}\ .
\label{I3}
\eq
 
Thus we conclude that
 \bq
\wt I_F\,\to\,0
\quad{\rm as}\quad
N\,\to\,\infty\ .
\label{I=0}
\eq
That is, in the long evolved state, the invariants $I_F$ are exclusively
contained in the coherent structure $(a_s,\ps_s)$, which therefore inherits the
exact magnetic topology of the initial state.  On the contrary, the energy and
the cross topology invariants, due to their linear divergence at $N\to\infty$,
are shared between the coherent structure and the fluctuations.  
 
Analogously to conserved quantities, we can estimate the mean square
norm of the fluctuations:
 \bq
\left<\bff|\bff\right>\equiv
\int(\wt a^2 + \wt\ps^2)d^2\bfx=
\sum_m\frac{1}{\la_m}\le
\frac{E_0\cals}{2\pi(1-|\ga'|_\smax)}\,\frac{\ln N}{N}
\,\to\,0\ ,\quad{\rm as}\quad N\to\infty\ .
\label{NORM=0}
\eq
 
Thus the mean square amplitude of the fluctuations, measured in the magnetic
flux and the stream function, goes to zero in the continuum, or the long-time
limit $N\to\infty$.  It is emphasized that the amplitude of {\em all}, not only
higher, fluctuation modes goes to zero.
 
The assumption of $\wt a\ll a_s,\;\wt\ps\ll\ps_s$ was indeed necessary for the
quadratic expansion of the probability functional $P[a,\ps]$ near the
equilibrium leading to the well-behaved Gaussian distribution.  In fact, the
small amplitude of $\wt a$ and $\wt\ps$ is not sufficient to apply the Gibbs
formalism, because the fluctuations in $\bfna\wt a$ and $\bfna\wt\ps$, which
enter the integrals of energy ({3}) and cross topology ({7}), are found
to be not small.  Fortunately, the quadratic expansion of Eqs.~({3}) and
({7}) is also valid, because the energy and the cross topology invariants
are {\em themselves quadratic (bilinear) with respect to the derivatives}
$\nabla a$ and $\nabla \psi$. 
This fortune is not extended to many other systems, most notably the
two dimensional Euler equation, where the resulting non-Gaussianity of the
fluctuations makes it difficult to draw any quantitative conclusions based on
the  Gibbs ensemble (see Appendix~D).

\section{Iso-topological relaxation and topological attractors}
\label{sec:relaxation}

Now the formal procedure of varying the integral of entropy ({26})
[which leads to the equilibrium (28)---(29)] can be rendered more
physical sense.  Namely, the coherent structure minimizes the energy subject to
the conservation of all magnetic topology invariants ($I_F,\ \forall F$) and one
of the cross topology invariants ($J_\ga$) of the initial condition:
 \bq
\min\,E[a,\ps]\bigg|_{
I_F[a]=\const,\ \forall F,\ {\rm and}\ J_\ga[a,\ps]=\const}\ .
\label{isotop}
 \eq
The term ``iso-topological relaxation'' describes a process whereby
this minimization may be achieved.  As we are now interested in the coherent
(coarse-grained) part of the MHD system, and the fluctuations are set aside,
the discussed relaxation is no longer Hamiltonian and resembles that
occurring in dissipative systems.  In fact, the usual dissipation in
macroscopic systems also originates from the purely Hamiltonian
molecular dynamics, where one is not interested in the microscopic
degrees of freedom.
Due to the seemingly dissipative nature of the iso-topological
relaxation, the relaxed state may be considered as an attractor.  The
qualitative arguments developed in this section predict the appearance
of the relaxed state without solving the complicated nonlinear problem
of the reconstruction of the functions $\ga(a)$ and $\be(a)$ from the
initial state, a task which must be complete for a quantitative
prediction and appears to be feasible only numerically.
 
Examine the equation of the coherent structure (29), or its variational
form (55).  Introduce the modified magnetic flux function by the
ansatz
 \bq
a'=\Ga(a)\ ,
\label{ansatz}
\eq
where the function $\Ga(a)$ is defined by Eq.~(30) and is in principle
known from the initial condition.  We note that the conservation of the 
magnetic topology invariants $I_F$ for the field $a$ is equivalent to the
conservation of those for the modified field $a'$.  Then Eq.~(29) can
be interpreted as an equilibrium condition for the modified magnetic field $a'$
with no fluid flow.  In other words, the problem reduces to the incompressible,
iso-topological minimization of the modified magnetic energy
 \bq
E'_m=\frac{1}{2}\int(\bfna a')^2\,d^2\bfx\ ,
\label{45}
\eq
starting from the specified initial condition $a'_0(\bfx)=\Ga(a_0(\bfx))$. 
Upon minimizing Eq.~(57) subject to the iso-topological variation $\de
a'=\{\mu,a'\}$ with arbitrary $\mu(x,y)$ results in $\{\bfna^2a',a'\}=0$,
implying a functional dependence between the modified magnetic flux and the
modified current.
 
Once the relaxed state $a'_\infty(\bfx)$ is found, the coherent structure of
the original magnetic field is recovered by inverting Eq.~(56):
 \bq
a_s(\bfx)=\Ga^{-1}(a'_\infty(\bfx))\ .
\label{relaxed}
\eq
Then the stream function of the fluid flow in the coherent structure is given by
 \bq
\ps_s(\bfx)=-\ga(a_s(\bfx))\ .
\label{relaxed_psi}
\eq
 
The kind of relaxation undergone by the modified field $a'$ will take
place in an incompressible, viscous fluid with an ideal conductivity,
where the viscosity damps down the fluid motion.  It is well known that
such an iso-topological relaxation may not be attainable in the class
of smooth magnetic fields \cite {Syrovatskii71,Arnold74,Moffatt90}.  In
two dimensions, these are the saddle ($x$) points of the initial
magnetic field that lead to singularities---current sheets---in the
relaxed field.  It is important to note that the location and the shape
of the current sheet is not locally determined by the $x$ point alone,
but rather depends on the shape of the separatrix coming through the
$x$ point.  The qualitative arguments of Ref.~\cite {Gruzinov93a} show
that each initial magnetic separatrix, in course of the iso-topological
relaxation, turns into a characteristic structure with a current
sheet---the {\em asymptotic separatrix structure\/} shown in Fig.~1.  
\begin{figure}
\centerline{
  \psfig{figure=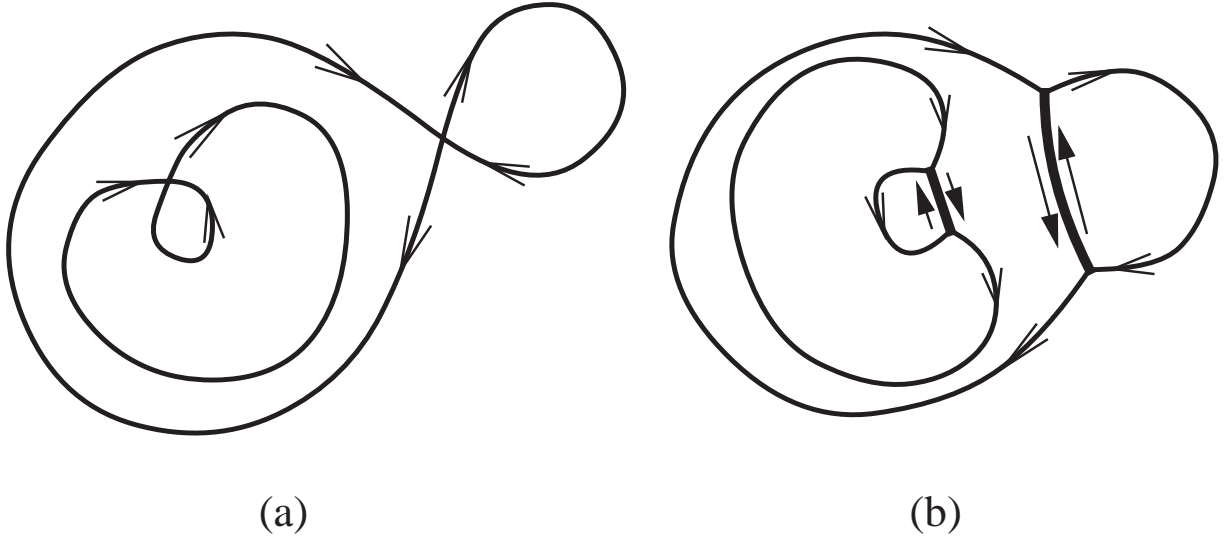,width=6.5in}
}
\caption{
Asymptotic separatrix structures resulting from iso-topological
relaxation.  Topologically nontrivial initial state (a) leads to the
formation of a relaxed state (b) with current sheets (shown bold).  The
arrows indicate the direction of the magnetic field.
}
\end{figure}
The orientation of the current sheet is such as to lie within a ``figure
eight'' separatrix and to border the outside of an ``inside-out figure
eight.'' 
 
It appears that the final state $a'_\infty$ of the iso-topological
relaxation is uniquely determined by the initial state $a'_0$.  The
same is true of the corresponding magnetic fluxes $a_s$ and $a_0$, once
the function $\ga(a)$ [and thereby $\Ga(a)$] is known.  Even without
any information about $\ga(a)$ the appearance of the relaxed state is
well understood qualitatively through the above construct, because
applying a function to $a$ does not change the geometry of magnetic
field lines.

\subsection{Comparison with numerical data}
\label{sec:numerical}
 
The computation of the long-time evolution of nearly ideal 2D MHD
turbulence reported by Biskamp and Welter \cite {BW89} clearly shows
current sheets terminating at $Y$ points, which are characteristic of
the asymptotic separatrix structures, although the reconnection due to
finite magnetic (hyper)diffusivity smears out the individual topology
of separatrices.
 
The spatial distribution of the fluctuations $(\wt a,\wt\ps)$ is
determined by the eigenfunctions of operator ({34}).  The potential of
this operator involves second spatial derivatives of $a_s$ (through the
term $\be^*$).  The singularities of $j_s=-\bfna^2a_s$ are
delta-function singularities at current sheets.  This must lead to the
localization of the wavefunctions (the fluctuations) near the potential
wells (the current sheets) where $\ga''(a_s)\bfna^2\ga(a_s)<0$.  This
kind of localization of the microscopic turbulence near the current
sheets is indeed observed in the computation of Ref.~\cite {BW89}.
Earlier simulations of turbulent magnetic reconnection \cite
{ML86,MM81} also confirm this picture.

\subsection{Iso-topological relaxation and magnetic reconnection}
\label{sec:dissipation}
 
\label{loc7}
So far we were mostly concerned with the ideal model of two-dimensional
MHD, and the question is in order as to the evolution of a more
realistic dissipative system involving finite electrical resistivity 
and fluid viscosity.
In general, this is a very difficult problem, because no
straightforward perturbation theory can be built for small coefficients
appearing in front of higher derivatives in the equations.  We
therefore restrict ourselves to the qualitative analysis of the role of
small dissipation.

If the dissipation is small, the system behavior clearly must resemble,
up to a certain point, the prediction of the ideal MHD theory.  The
deviation of a weakly dissipative evolution from the ideal behavior is
always a matter of time of the evolution.  In order to neglect the
effects of dissipation in MHD turbulence relaxation, not only must the
resistivity $\et$ and the viscosity $\nu$ be small but also the length
scales should be sufficiently large.  The ideal MHD evolution discussed
above does lead to the formation of small scale structures which
trigger, in the long run, the strong effects of the weak dissipation.
 
If the initial state is smooth, the small-scale structures do not
appear at ones; it takes several nonlinear (eddy turnover) times for
the small scales to show up.  In the meantime, the system evolves
towards, however not quite attains, the ideal statistical equilibrium.
In fact, the principal manifestation of approaching the statistical
equilibrium is the separation of scales into long-wavelength coherent
structures and short-wavelength fluctuations.  It is reasonable to
assume that this separation of scales is not only necessary, but also
{\em sufficient\/} for the statistical equilibrium to set in.  Then, by
the time when the initially small dissipation becomes important, the
coherent part of the turbulent field is essentially built by the
statistical mechanics of ideal MHD turbulence.  The smaller the
dissipation, the shorter scales are allowed to evolve in the
Hamiltonian fashion, and therefore the closer the attained shape of the
coherent structures to the exact predictions of the Gibbs-ensemble
theory.

The time scale $\ta^*$ specifying the crossover from the ideal regime to
the dissipative regime is certainly much shorter than the diffusive
time $\ta_\et\sim(k_s^2c^2\et)^{-1}$ and may not be very long compared
to the characteristic nonlinear time $\ta_A$.  Numerical 
results \cite{MSMOM91,MMSMO92} indicating the enstrophy decay in 2D
fluid in just a few eddy turnover times suggest that $\ta^*/\ta_A$ is a
small power or even logarithm of the large Reynolds number.  The fast
crossover to the dissipative regime directly indicates the fast
production of small scales and, therefore, the equally fast approaching
to the statistical equilibrium.

After the approximate equilibrium is set in, the dissipation takes over
and the small-scale fluctuations are significantly damped over several
times $\ta^*$, whereas the coherent structures remain little affected,
at least in the case when these structures involve no singularities.

If the initial magnetic field has $x$ points, the coherent structure
will develop current sheets.  The coherent structure will then undergo
fast magnetic reconnection.  The reconnection occurs in a
characteristic time $\ta_r$ much longer than the \Alfvenic\ time
$\ta_A$, if the magnetic Reynolds number $R_m=\ta_\et/\ta_A$ is large.
By different models, $\ta_r/\ta_A$ ranges from $R_m^{1/2}$
\cite {Sweet58,Parker57} to $(\ln R_m)^{p},\ p>0$ \cite 
{Petschek64,Sonnerup70}, although the former (Sweet-Parker) model
appears to be more typical \cite {Biskamp85}.

So the ideal MHD turbulence theory describes the early,
$t<\min(\ta^*,\ta_r)$, iso-topological stage of the turbulent MHD
relaxation and predicts the appearance of the coherent structures
entering the later stages where magnetic reconnection and/or viscosity
play the dominant role.  Even then, some topological invariants survive
better than others, also providing useful variational tools for the
prediction of fully relaxed \cite 
{Taylor74,Kadomtsev75,Taylor86,MS87,AT91} 
\label{loc8}
or selective-decay \cite {MSMOM91,MMSMO92} states.  
 
Our theory can be used to qualitatively describe the relatively early
stage, $\ta_A\ll t\ll(\ta_A\ta_\et)^{1/2}$, of the nonlinear kink
tearing mode in a tokamak, where two dimensional MHD models are
commonly used for helically symmetric magnetic perturbations \cite
{KP73,Kadomtsev75,RMSW76,Waelbroeck89}.  The kink tearing is
accompanied by changes in magnetic topology.  First, an $x$ point in
the ``auxiliary magnetic field'' $\bfB_*=\bfB-q\bfB_\th$ is created
near the linearly unstable $q=1$ surface.  Then the resulting
``magnetic bubble'' is pushed to the exterior of the plasma column by
essentially ideal MHD motions.  This process is likely to be of
turbulent nature and, until very small scales are generated, the ideal
turbulent relaxation will proceed in the direction of forming a
coherent structure with a current sheet corresponding to the initial
$x$ point, as suggested by the Gibbs statistics.  This stage of
evolution may be pretty long, as the magnetic Reynolds number in
tokamaks can be of order $10^6$ and more.  Later on, magnetic
reconnection via the current sheet \cite {Biskamp85} will occur at a
characteristic time of order $(\ta_A\ta_\et)^{1/2}$.  Dynamically, the
reconnection develops through a sequence of singular MHD equilibria
with the same local helicity
\cite {Kadomtsev75}, as analytically described by Waelbroeck
\cite {Waelbroeck89}.  The first of the sequence of these current-sheet
equilibria can be interpreted as the asymptotic separatrix structure
arising from the initial state via the iso-topological turbulent
relaxation.

\section{Summary and conclusion}
\label{sec:conclusion}
 
The main result of this paper lies in working out the Gibbs statistics
for a Hamiltonian PDE system with an infinity of constants of the
motion.  This formalism was demonstrated in the example of two
dimensional magnetohydrodynamics but can be carried over to other
systems.  We review again the principal steps of our approach in terms
of a general nonlinear Hamiltonian system describing the fields
$\bfps(\bfx,t)$ and having a finite or an infinite number of invariants
$\bfI[\bfps]=I_1[\bfps],I_2[\bfps],\ldots$.  Here it does not matter
what these invariants are; one can think of $I_1$ as the energy and of
the rest as topological invariants.
 \renewcommand{\labelenumi}{(\alph{enumi})} 
 \begin{enumerate}
 \item 
The solution to the underlying nonlinear system is sought in the form
$\bfps(\bfx,t)=\bfps_s(\bfx)+\wt\bfps(\bfx,t)$, where $\bfps_s(\bfx)$
is yet unspecified stationary, Lyapunov stable solution (coherent
structure).  We then anticipate that the amplitude of the fluctuation
field $\wt\bfps(\bfx,t)$ is going to be small and hence the exact
integrals of motion can be expanded about the coherent structure
$\bfps_s$ up to quadratic terms: $\bfI=\bfI_s+\wt\bfI$.
 \item 
The Gibbs ensemble is introduced in the fluctuation space in the
standard form of the exponential of a linear combination of all
invariants, $P[\wt\bfps]=\exp(-\bfal\cdot\wt\bfI)$, where
$\bfal=\al_1,\al_2,\ldots$ are the reciprocal temperatures to be
determined from the initial state.  Having an infinity of Casimirs,
which depend on an arbitrary function, is not an obstacle, because any
linear combination of the Casimirs is again one of them.  In order to
have non-diverging fluctuations, we exercise our right to suitably
choose the coherent structure $\bfps_s$.  Namely, $\bfps_s(\bfx)$ is
required to minimize the linear combination $\bfal\cdot\bfI[\bfps]$ of
the invariants, where the stationarity of the resulting state is
ensured by the Arnold variational principle.  Then $\bfal\cdot\wt\bfI$
is a positive definite quadratic form, and the Gibbs distribution of
the fluctuations is a Gaussian distribution.  From now on, the standard
Boltzmann-Gibbs statistics is applied in a straightforward way, at
least for a finite dimensional approximation using $N$ eigenmode
amplitudes $f_m$ satisfying the Liouville theorem.
 \item 
The eigenmodes are introduced such as to diagonalize the Gibbs
exponential, $\bfal\cdot\wt\bfI=(\al_1/2)\sum_m\la_mf_m^2$.  Then
averages can be computed in the conservation laws,
$\bfI_0=\bfI_s+\left<\wt\bfI[\wt\bfps]\right>$, in order to infer the
equations for the temperatures.
 \item 
The fluctuations' share of the invariants,
$\left<\wt\bfI[\wt\bfps]\right>=\sum_m1/(2\al_1\la_m)\p^2\wt\bfI/\p
f_m^2$, when expanded in the eigenmodes, turns out to diverge as
$N\to\infty$, unless the temperatures $1/\al_m$ are let to zero (even
then the temperature ratios remain finite and keep useful information).
This is the ``ultraviolet catastrophe.'' The regularization of this
divergence requires the reciprocal temperatures to also diverge, e.g.,
$\al_1(N)\propto N$.
 \item 
If the square norm of the fluctuations
$\left<\wt\bfps|\wt\bfps\right>=\sum_m(\al_1\la_m)^{-1}$ diverges at
constant temperatures slower (e.g., logarithmically) than the fastest
diverging invariant (say, the energy $I_1$), then the average norm goes
to zero as $N\to\infty$.  This is the crucial point, which justifies
the assumption of the small amplitude necessary for the Gaussianity of
the fluctuations in the given representation.  If this condition is not
fulfilled, one can always pick other variables involving lower order of
derivatives, such as $\bfps'=\bfna^{-2}\bfps$, and repeat the above
steps.  However, the exact Gaussianity of the fluctuations requires
another important property of the integrals of motion, which is
independent of the variables used.  Namely, {\em each invariant must be
not more than quadratic in the highest-order-derivative variables}.
Then the quadratic expansion will be also valid even for those (fastest
diverging) invariants, whose fluctuations are finite.  This property
holds for 2D MHD, but it does not for 2D Euler turbulence or
Vlasov-Poisson system (Appendix~D). 
 \item 
If, in addition, there are invariants (such as magnetic topology
invariants) diverging slower than the fastest diverging integral of
motion, then the average fluctuation's share of those invariants
vanishes as $N\to\infty$.  The presence of such invariants simplifies
the analysis of the coherent structure.
 \end{enumerate}
 \renewcommand{\labelenumi}{\arabic{enumi}.}  
 
We use the above steps to study the relaxation of ideal two dimensional
MHD turbulence, where both infinite sets of topological invariants,
magnetic ({6}) and cross ({7}), are incorporated.  We show that
accounting for all topological invariants leads to the prediction that
the long evolved MHD turbulent state consist of a coherent structure
(the most probable state) and a small-amplitude, small-scale Gaussian
turbulence (the fluctuations).  The fluctuations are small if measured
in terms of the magnetic vector potential $\wt a$ and the flow stream
function $\wt\ps$.  The fluctuations in the magnetic field $\wt\bfB$
and the fluid velocity $\wt\bfv$ are of the same order as in the
coherent structure.  The fluctuation current $\wt\bfj$ and vorticity
$\wt\om$ are infinite in the long time limit.
 
We find that in 2D ideal MHD turbulence the coherent structure has the
same magnetic topology as the initial state, while energy and cross
topology are shared between the coherent structure and the
fluctuations.  Therefore, for a sufficiently wide class of initial
conditions having the same topological invariants, the final coherent
state is the same, whereas the fluctuations, when measured by the
standard norm (54), become asymptotically ``invisible.'' In this sense,
the coherent structures emerging from the turbulent MHD relaxation can
be regarded as ``topological attractors,'' even though the underlying
dynamics is perfectly Hamiltonian.  (The theorem of the absence of
attractors in Hamiltonian systems is not valid for infinite dimensional
PDE systems.)
 
The presence of the fluctuations on the top of the coherent structure
is conceptually important even though the amplitude of these
fluctuations goes to zero in the long time limit: the fluctuations
appear as the storage of the ``lost'' integrals of motion, if only the
most probable state is compared with the initial state.  This explains
the well-known result (cf.~\cite {Miller90}), that the topological
invariants of the coherent vortex emerging from 2D Euler turbulence,
are different from those of the initial state.  In 2D
magnetohydrodynamics the role of the initial topology is more
important.  In Appendix~D we discuss the application of the
Gibbs-ensemble formalism to the two dimensional Euler equation.
 
We formulate the variational principle of iso-topological relaxation,
which allows us to predict the shape of the coherent structure for the
given initial state.  We show how the problem of the ideal MHD
relaxation with plasma flow is reduced to the viscous relaxation of
magnetic field with no flow in the final state.  The numerical results
suggest that the asymptotic separatrix structures with current sheets
are indeed observed during the turbulent relaxation.  It appears that
these structures are the route to reconnection in the nonlinear kink
tearing mode in tokamaks.
 
Many problems of MHD turbulence remain, most notably the role of small
dissipation.  As discussed in Sec.~\ref{sec:dissipation}, this is the
dynamics of producing small scales which determines when and how the
dissipative processes become important.  In order to study the
phenomena of crossover from the ideal to the dissipative turbulent
relaxation, the nonequilibrium dynamics of the ideal relaxation must be
worked out.
It appears that the formalism of the weak turbulence theory
\cite{FS91,ZLF92,Pomeau92b} can be appropriately suited for 
Eq.~(\ref{f-dynamics}) in order to study the nonequilibrium 
statistics of 2D MHD turbulence.  
\label{loc9}
However, the important role of the
ideal Gibbs turbulence for weakly dissipative systems is found in that
the ideal turbulence forms predictable coherent structures, which enter
the later, dissipative stages of the turbulent evolution.

The comparison of the MHD and the Euler turbulence prompts us to
distinguish between three kinds of advected fields.  The first kind is
passive field, such as the concentration of a dye or the temperature
that do not affect the advecting velocity field.  Passive fields tend
to become spatially uniform due to turbulent diffusion.  The second
kind is active field, such as the vorticity in Euler fluid, which does
affect the velocity field but whose lines or contours can be
indefinitely stretched at no significant energy price.  The active
fields therefore tend to self-organize assuming topologically simple
structures, like monopole vortices, whose topology is different from
that of the initial state.  The third kind can be referred to as
``reactive field,'' such as the magnetic flux frozen into an ideally
conducting fluid.  Stretching of magnetic field lines is energetically
expensive and cannot last indefinitely.  The topology of the reactive
field is therefore much more robust than that of passive or active
fields, and the self-organization can lead to nontrivial coherent
structures with singularities (current sheets).

\subsection*{Acknowledgments}
 
We wish to thank V.~V. Yankov, P.~J. Morrison, F.~L. Waelbroeck, F.
Porcelli, P.~H. Diamond, G.~E. Falkovich, and J.~B. Taylor for
stimulating discussions.  This work was partially supported by the
U.S.~Department of Energy under Contracts No.~DE-FGO3-88ER53275 and
DE-FG05-80ET53088.
 
\clearpage
 
\appendix   
\renewcommand{\theequation}{\thesection.\arabic{equation}}

\section*{APPENDICES}   

\setcounter{equation}{0}
\section{Extremal properties and stability of MHD and Euler
 equilibria:  Positive and negative temperatures} 
\label{A}
 
Lyapunov stability of equilibria in conservative systems depends on the
extremal properties of their invariants.  The strongest form of the
relevant Lyapunov theorem states that if a state realizes a {\em
conditional\/} nondegenerate extremum (that is, a minimum or a maximum)
of one of the invariants subject to the conservation of one or more of
other invariants, then this state is stable.  It is emphasized that the
above is the sufficient criterion for nonlinear stability with respect to
{\em any\/} sufficiently small perturbations.  Originally formulated for
finite dimensional systems, the Lyapunov theorem is extended to PDE
systems on a case-by-case basis and depends on the functional norm used
to specify what ``small'' means \cite {Arnold78}.
 
The existence of such extremal, and therefore stable, states can be
detected by inspecting various inequalities involving integrals of
motion.  As an example, consider 2D MHD states with the fixed cross
helicity ({8}),
 \bq
\int\bfna a\cdot\bfna\ps\,d^2\bfx=J_1\ .
\label{J_1}
\eq
Then the energy ({3}) is clearly bounded from below,
 \bq
E\ge|J_1|\ ,
\label{E<J_1}
\eq
which indicates the existence of a minimum-energy state subject to the
conservation of the topological invariants.  An example of such a state
is any circular magnetic configuration $a=a(r)$ with no fluid flow,
$\ps=0$.  This configuration assumes the absolute minimum of energy
with respect to all neighboring iso-topological MHD states.  Indeed,
the magnetic force is the tension of magnetic field lines, which tend
to shrink while preserving the area inside them.  The smallest
perimeter at fixed area is assumed by a circle.
 
A formal proof, which can be extended to more interesting geometries,
can be given in terms of the iso-topological energy variation
(15)---(16).  The first variation is vanishing at an equilibrium state,
and the second is given by the general form
 \bqy
\hspace*{-15pt}\de^2E_m &=& \int \left[ -\{\mu,a\}\{\mu,j\} + 
\left(\bfna\{\mu,a\}\right)^2 \right] d^2\bfx\ ,
\label{16}
\\
\hspace*{-15pt}\de^2E_f &=& \int
\left[ (\bfna\de\ps)^2-\{\mu,\ps\}\{\mu,\om\}-\{\mu,\ps\}\{\nu,a\}-
\{\mu,a\}\{\nu,\ps\} \right] d^2\bfx\ ,
\label{17}
\eqy
for the magnetic and the fluid kinetic energy, respectively.
 
Upon letting $a=a(r)$ and Fourier expanding the
azimuthal-angle-periodic displacement function
$\mu(r,\th)=\sum\mu_m(r)\exp(im\th)$, several integrations by parts
(presuming that $\mu$ goes to zero sufficiently fast for both $r\to0$
and $r\to\infty$) yield
 \bq
\de^2E_m = \sum_{m=-\infty}^{\infty}m^2\int_0^{\infty}
\frac{dr}{r^3}\left(\frac{da}{dr}\right)^2
\left(m^2|\mu_m(r)|^2+|\mu_m(r)-r\mu_m'(r)|^2\right) > 0\ .
\label{46}
\eq
Equation (A.4) also implies $\de^2E_f>0$ for $\ps=0$.  That is, any
circular magnetic field without fluid flow assumes the conditional {\em
minimum} of energy subject to the conservation of topological
invariants and is therefore Lyapunov stable.  Adding a small fluid
velocity along the circular magnetic field lines will not change the
stability of such an equilibrium.
 
It is easy to see that there is no upper bound on the MHD energy, even
though all the topological invariants ({6})-({7}) are fixed: by an
incompressible convolution of the magnetic field lines the magnetic
energy can be made arbitrarily large.  
 
The kinetic fluid energy can be represented in the form of
``electrostatic'' energy of charges with the density $\om$,
 \bq
E_f = \frac{1}{2}\int \ps\om \,d^2\bfx = \frac{1}{2}
\int\om(\bfx_1)\om(\bfx_2)G(\bfx_1,\bfx_2)\,d^2\bfx_1d^2\bfx_2\ ,
\label{charges}
\eq
where the Green function $G$\ \ [$=-(1/2\pi)\ln|\bfx_1-\bfx_2|$ for an
infinite domain] plays the role of the interaction potential.  The
total charge, $J_0=\int\om d^2\bfx$, is conserved, and the charges are
allowed to redistribute only along the magnetic field lines [in MHD:
invariants ({7})] or by means of incompressible interchanges [in Euler
equation: integrals (D.3)].
 
The extremal properties of two dimensional Euler fluid are different
from those of MHD in that the energy has also an upper bound at fixed
topological invariants.  According to the Schwarz inequality, we have
 \bq
E_f^2 \le \frac{1}{4}
\int \om^2(\bfx_1)\om^2(\bfx_2)\, d^2\bfx_1d^2\bfx_2
\int G^2(\bfx_1,\bfx_2)\,         d^2\bfx_1d^2\bfx_2=\const\,I_2^2\ ,
\label{schwarz}
\eq
where $I_2=\int\om^2d^2\bfx$ is the conserved fluid enstrophy, and we
have used the quadratic integrability of the Green function, which has
only logarithmic singularity at $\bfx_1=\bfx_2$.  By collecting the
most intensive charges closest to each other---that is, forming a
circular vortex with a monotonically decreasing vorticity of the same
sign, we construct the maximum-energy state \cite {PY89}.  Such
vortices play an important role in 2D Euler turbulence \cite
{CMcWPWY91}.  Conversely, a monotonically increasing vorticity in a
circular domain assumes the minimum of energy under fixed topological
invariants.
 
Adding even a small amount of magnetic field to the maximum-energy
stable vortices will drive them unstable, because the negative-energy
waves perturbing the vortices will dump their energy into the magnetic
field via magnetic dynamo and thereby grow in amplitude.  The resulting
turbulent relaxation will lead to other, minimum-energy equilibria,
which may be non-circular due to the competition of the elastic tension
of magnetic field lines with the repulsion of the ``vorticity charges''
threaded onto these lines.
 
As discussed in Sec.~\ref{sec:fluctuations}, the Gibbs distribution of
fluctuations about an equilibrium coherent structure has the form of 
 \bq
\exp[-\al\de^2(E+{\rm Topological\ Invariants})/2]\ , 
\label{general_gibbs}
\eq
where $\al=1/T_E$ is the reciprocal temperature conjugate to energy and
$\de^2$ the second (unconstrained) variation specifying the fluctuations.  In
order for the fluctuations to be finite, the temperature must be positive for
minimum-energy states, as it is in ordinary world or in 2D MHD, and negative for
maximum-energy states, as is possible in 2D ideal fluid 
\cite {Onsager49,MJ73,ET74}

\setcounter{equation}{0}
\section{Liouvillianity of dynamical variables}
\label{B}
 
The probability distribution functional ({38}) is meant to describe the
statistical properties of the fluctuations.  The dynamics governing the
amplitudes $f_m(t)$ can be written by substituting the eigenmode
decomposition (32) into Eqs.~({1}) and ({2}).  Then, using the
orthonormal properties of the eigenmodes, we obtain
 \bq
\dot f_m=\sum_nH_{mn}f_n+\sum_{np}H_{mnp}f_nf_p\ ,
\label{f-dynamics}
\eq
where
\bqy
H_{mn}&=&\int
[
  a_m\{\ps_n,a_s\}+a_m\{\ps_s,a_n\}+
  \nonumber\\
  \quad&&
  \om_m\{\ps_n,\om_s\}+\om_m\{\ps_s,\om_n\}+\om_m\{j_n,a_s\}+\om_m\{j_s,a_n\}
]
d^2\bfx\ ,
\label{Hmn}\\
H_{mnp}&=&\int\left[
a_m\{\ps_n,a_p\}+\om_m\{\ps_n,\om_p\}+\om_m\{j_n,a_p\}
\right]d^2\bfx\ ,
\label{Hmnp}
\eqy
the indices $m,n$, and $p$ refer to the eigenmodes and $s$ to the coherent
structure.  The divergence of the phase-space flow defined by
Eq.~(B.1),
 \bq
\sum_m\p\dot f_m/\p f_m=
\sum_mH_{mm}+\sum_{mn}(H_{mmn}+H_{mnm})f_n\ ,
\label{B.4}
\eq
must be zero for Liouvillian variables.  This is generally not true even
for the first term on the RHS of Eq.~({B.4}).  However, a linear
change to new variables can kill the zero-order term in
Eq.~({B.4}).  An example of new variables is given by the eigenmode
amplitudes of another linear operator,
 \bq
W'\left(\begin{array}{l}
         \wt a\\
         \wt\om
       \end{array}\right)=
\left(\begin{array}{ll}
         (\be^*-\bfna^2) & \quad\ga'\\
         \ga'            & -\bfna^{-2}
\end{array}\right)
\left(\begin{array}{l}
         \wt a\\
         \wt\om
       \end{array}\right)
\equiv
\left(\begin{array}{c}
         \be^*\wt a+\wt j+\ga'\wt\om\\
         \ga'\wt a + \wt\ps
\end{array}\right)\ ,                          
\label{W'}
\eq
such that the fluctuation part of $E+I_\be+J_\ga$ can be formally written as 
 \bq
\left<\begin{array}{l|}
         \wt a \\
         \wt\om  
\end{array}\right.
W'
\left.\begin{array}{|l}
         \wt a \\
         \wt\om  
\end{array}\right>\ .
\label{flS}
\eq
Then, in terms of the new eigenmodes, the equation of motion can be
written exactly as Eqs.~(B.1), (B.2), and (B.3); however, the equations
for the eigenfunctions $a_m$ and $\om_m$ are different from those of
the old operator $W$ ({34}).  The usefulness of the new variables is
that the zero-order term $H_{mm}$ in Eq.~({B.4}) identically vanishes
after integrating by parts for each $m$, whereas the linear term,
 \bq
H_{mmn}+H_{mnm}=
\int\left[
\ga'\{(a_m^2+\om_m^2)/2,a_n\}+\ga'\{\om_m,a_m\om_n\}+\be^*\{\om_m,a_ma_n\}
\right]d^2\bfx\ ,
\label{H+H}
\eq
generally persists unless there is no inhomogeneity due to the coherent
structure. Nevertheless, the remaining linear term in Eq.~({B.4})
vanishes as the amplitude goes to zero.  Thus the new variables $f_m$
are Liouvillian only in the limit of vanishing amplitudes, which is
also the assumption lying behind the Gaussian distribution ({38}).  The
validity of this assumption is confirmed in Sec.~\ref{sec:partition}.
 
As the change from the old to the new variables is a {\em linear\/}
transformation, the non-Liouvillianity of the old variables only leads
to a {\em constant} weighting factor---the Jacobian---in the
probability ({38}), which does not affect the averaging.  Hence we may
write $\prod_mdf_m$ in the functional integrals in ({22}) instead of
$\cald a\cald\ps$.

\setcounter{equation}{0}
\section{The spectrum of the eigenmodes}
\label{C}
 
The eigenmode equation for the operator ({34}) can be written
 \bqy
(\la_m-\be^*+\bfna^2)a_m&=&-\ga'\bfna^2\ps_m\ ,
\label{eigen1}\\
\left[-\ga'\bfna^2-2\bfna\ga'\bfna-(\bfna^2\ga')\right]a_m&=&
(\la_m+\bfna^2)\ps_m\ .
\label{eigen2}
\eqy
We are interested in the high-mode (WKB) limit $m\to\infty$, where we can
replace $\bfna$ acting on $a_m$ and $\ps_m$ by $i\bfk$.  Then we get the
quadratic equation for the eigenvalues,
 \bq
\la_m^2-(2\bfk^2+\be^*)\la_m+\bfk^4(1-\ga'^2)+i\bfk^2\bfk\cdot\bfna\ga'^2+
\bfk^2(\be^*+\ga'\bfna^2\ga')=0\ ,
\label{quadratic_equation}
\eq
having the solutions
 \bq
\la^\pm_m=\bfk^2(1\pm\ga')+\frac{\be^*}{2}\mp
\frac{i\bfk\cdot\bfna\ga'}{\ga'\bfk^2}+\calo(k^{-2})\ .
\label{eigenvalues}
\eq
The corresponding eigenfunctions satisfy
 \bq
\frac{\ps_m}{a_m}=\frac{\la_m-\bfk^2-\be^*}{\ga'\bfk^2}=
\pm1-\frac{\be^*}{2\ga'\bfk^2}+\calo(k^{-3})\ .
\label{eigenfunctions1}
\eq
The ordering of the modes with number can be figured out by substituting the
zero-order eigenfunction ratio (C.5) into Eq.~(C.1):
 \bq
\left(\frac{\be^*}{1\pm\ga'}-\bfna^2\right)a_m=\frac{\la_m}{1\pm\ga'}a\ .
\label{quasischrodinger}
\eq
If $\ga'$ were constant, Eq.~(C.6) would be a standard
Schr\"odinger equation with the Bohr-Sommerfeld quasiclassical energy levels
$\la_m^\pm=(1\pm\ga')4\pi m/\cals$ and the wavenumber (43).  In
the case of inhomogeneous $\ga'$ we arrive at formulas ({40}) and
(41).

\setcounter{equation}{0}
\section{Gibbs statistics of two dimensional Euler turbulence}
\label{D}
 
Although the Euler equation,
 \bq
\p_t\om = \{\ps,\om\}\ ,\quad\om\equiv-\bfna^2\ps\ ,
\label{euler}
\eq
is a special case of the MHD system ({1})-({2}) at zero magnetic field,
the properties of two dimensional Euler turbulence are drastically different
from those of 2D MHD turbulence.  This is primarily due to the different
structure of the integrals of motion, which for the Euler fluid are energy
 \bq 
E=\frac{1}{2}\int(\bfna\ps)^2\,d^2\bfx\ ,
\label{eulerE}
\eq 
and vorticity topology invariants
 \bq
I_F = \int F(\om)\,d^2\bfx\ .
\label{eulerI}
\eq
 
The formal Gibbs ensemble
 \bq
P[\ps]=\exp[-\al(E+I_\be)]\simeq 
P[\ps_s]\exp
\left(
-\frac{\al}{2}\left<\wt\ps|W|\wt\ps\right>
\right)
\label{eulerP}
\eq
can be approximately written in terms of the eigenmodes of the self-adjoint
operator
 \bq
W=-\bfna^2+\bfna^2\be''(\ps_s)\bfna^2\ ,
\label{eulerW}
\eq
whose quasiclassical eigenvalues behave as
 \bq
\la_\bfk\simeq\be''\bfk^4\ ,\quad
\bfk_m^2\simeq\frac{4\pi m}{\cals}\ ,\quad m\gg k_s^2\cals. 
\label{eulerLA}
\eq
 
Then the fluctuation energy,
 \bq
\left<\wt E\right>\simeq
\sum_m\frac{1}{2\al\la_m}\int(\bfna\wt\ps_m)^2\,d^2\bfx
\sim\sum_m\frac{\bfk_m^2}{\al\la_{\bfk_m}}\ ,
\label{eulerE1}
\eq
diverges with the number of modes logarithmically, whereas the topological
invariants,
 \bq
\left<\wt I_F\right>\simeq\sum_m\frac{1}{2\al\la_m}\int
F''(\om_s)(\bfna^2\wt\ps_m)^2\,d^2\bfx
\sim\sum_m\frac{\bfk_m^4}{\al\la_{\bfk_m}}\ , 
\label{eulerI1}
\eq
diverge linearly with $N$.  In this sense, we have something like the
equipartition of topology, rather than of energy, between the
fluctuation eigenmodes.  The square norm of the fluctuation stream
function, $\int\wt\ps^2d^2\bfx$, converges.
 
In accordance with the principles (a)---(f) set forth in
Sec.~\ref{sec:conclusion}, we conclude that in the long time limit the
fluctuations in the stream function $\wt\ps\to0$.  The fluctuation
energy (and hence the velocity $\wt\bfv$) also goes to zero, but a
finite share of the topological invariants gets into the fluctuations.

Hence the coherent structure emerging from 2D Euler turbulence has the
same energy, but different vorticity topology, as compared to the
initial condition.  This state of affairs makes it hard to predict the
appearance of the coherent structure from only the analysis of the
integrals of motion and requires dynamical considerations.  
 
The real difficulty, however, is that the highest-order-derivative
variable, the vorticity $\om$, enters the topological invariants (D.3)
in all powers, not only quadratically.  This makes the expansion of the
topological invariants (D.3) about the coherent structure inaccurate,
and hence the Euler turbulence {\em non-Gaussian}.  
\label{loc10}
As a result, the formally written Gibbs distribution (D.4) is useless
for making quantitative predictions, because, as mentioned earlier,
non-Wienerian probability functionals do not allow well-defined
averages.  This difficulty is present even if the validity of the Gibbs
ensemble (D.4) is not questioned.  We note, however, that this also
remains unclear whether or not it is possible to derive the standard
canonical ensemble from the first principles set forth in the
microcanonical ensemble for the case of non-additive integrals of
motion (see also discussion in Sec.~\ref{sec:ensemble}).  Thus the
``exact'' predictions based on the Gibbs statistics of 2D Euler
turbulence \cite {Miller90,RS91,MWC92} appear questionable, because the
result depends on the discretization used to solve functional
integrals, or, equivalently, on the relative size of the
``micro-cells'' used for a combinatorial treatment.  One example of
such a dependence is given by the discretization by means of point
vortices, where the equilibrium state depends on the (arbitrary chosen)
vortex strengths.  Another example, dealing with continuum fields, was
provided by Lynden-Bell \cite {Lynden-Bell67} in the context of the
Vlasov-Poisson model of a stellar system, where the equilibrium is
given by a combination of Maxwell exponentials depending on the
arbitrary partitioning of the phase space into micro-cells.  There, the
source of the ambiguity is the same as in the Euler equation: the
highest-order-derivative variable---the star distribution function
$f(\bfx,\bfv,t)$---enters the Casimirs $I_F[f]=\int F(f) d\bfx d\bfv$
not only quadratically.
 
The non-Gaussianity of 2D Euler turbulence suggests that the
Boltzmann-Gibbs formula (D.4) is not valid for such a system.
Nevertheless, this does not affect the general trend of splitting into
a stationary coherent structure and a small-scale, small-amplitude (in
appropriate variables) fluctuations.  The dynamics of such an evolution
can be also described in terms of structures.  If the initial state has
the length scale much less than the box size, the coherent part of the
turbulence evolves through weakly interacting nearly circular vortices,
whose number decreases with time due to the vortex merger \cite
{CMcWPWY91,CMcWPWY92,WMcW92}.  
\label{loc11}

\clearpage
\baselineskip12pt
\bibliography{../bib/turb,../bib/chaos,../bib/mbi}

\end{document}